\documentclass[a4paper,superscriptaddress,nofootinbib,12pt]{revtex4}
\usepackage{amsfonts}
\usepackage{amsmath}
\usepackage{amsthm}
\usepackage{bbm}
\usepackage{amssymb}
\usepackage{mathrsfs}
\usepackage{color}
\usepackage{hyperref}
\usepackage{stmaryrd}
\usepackage{mathrsfs}
\usepackage{dsfont}
\usepackage{wasysym}
\usepackage{eufrak}
\usepackage[a4paper]{geometry}
\usepackage{hyperref}
\usepackage{graphicx}
\def\be{\begin{equation}}
\def\ee{\end{equation}}

\def\d{\mathrm{d}}
\def\sk{\vskip .4cm}
\def\FF{\mathcal{F}}

\def\oR{{\bar{R}}}
\def\of{{\bar{\mathrm{f}}}}

\def\al{\alpha}
\def\be{\beta}
\def\st{\star}

\def\eq{\begin{equation}}
\def\la{\lambda}

\def\en{\end{equation}}

\def\DD{{\cal D}}
\def\bbk{\boldsymbol{k}}

\begin{document}

\title{Dispersion Relations in $\kappa$-Noncommutative Cosmology}
\author{Paolo Aschieri}
\email{paolo.aschieri@uniupo.it}
\affiliation{Dipartimento di Scienze e Innovazione Tecnologica,
Universit\`a del Piemonte Orientale, Viale T. Michel 11, 15121 Alessandria, Italy}
\affiliation{INFN, Sezione di Torino, via P. Giuria 1, 10125 Torino}
\affiliation{Arnold-Regge Centre, Torino, via P. Giuria 1, 10125 Torino}
\author{Andrzej Borowiec}
\email{andrzej.borowiec@uwr.edu.pl}
\affiliation{Institute of Theoretical Physics, University of Wroclaw, pl. M. Borna 9,
50-204 Wroclaw, Poland.}
\author{Anna Pacho{\l }}
\email{a.pachol@qmul.ac.uk}
\affiliation{Queen Mary, University of London, Mile End Rd.,
London E1 4NS, UK.\\ \phantom{P}}

\begin{abstract}
We study noncommutative deformations of the wave
equation in curved backgrounds and discuss the modification of the
dispersion relations due to noncommutativity combined with
curvature of spacetime. Our noncommutative differential geometry approach is based on  Drinfeld twist
deformation, and can be implemented for any twist and any curved
background. We discuss in detail the Jordanian twist --giving
$\kappa$-Minkowski spacetime in flat space-- in the presence of  a
Friedman-Lema\^{i}tre-Robertson-Walker (FLRW) cosmological background.

We obtain a new expression for the variation of the speed of light,  depending linearly on
the ratio $E_{ph}/E_{LV}$ (photon energy / Lorentz violation scale),
but also linearly on the cosmological time, the Hubble parameter and
inversely proportional to the scale factor.

\end{abstract}

\maketitle
\section{Introduction}
Recent years provided us with experimental confirmations of long existing theoretical models, from the Higgs boson discovery at the LHC to the gravitational waves detection by LIGO. The experiments providing evidence for quantum gravity are yet to be found. The difficulty with finding measurable implications of quantum gravitational models lies within the energy scale of the theory. 
Nevertheless, the physics at the Planck scale, describing
gravitational interactions at the quantum level, might be indirectly
investigated in the cosmological and astrophysical context. Approaches
to quantum gravity phenomenology  have
considered the possibility that the Planck scale quantum structure of
spacetime induces a modification of the wave dispersion relations
including dependence of the velocity of photons on their energy
\cite{ACEMN},\cite{ACEMNS},\cite{GAC-Majid}.
Gamma ray bursts (GRBs) are the brightest electromagnetic events in the
universe, they are emitted also at relatively high redshifts ($z\sim
9$) and offer an opportunity for testing dispersion relations associated with a Planck scale breaking of Lorentz symmetry, which even though very small may be amplified by the
cosmological distances. 
Indeed, data analysis related with the time delays of photons arriving from distant
GRBs  are consistent with the velocity of light
having a tiny dependence on its energy, cf.  \cite{constraints}
\cite{Liberati} and
references therein, and the recent studies \cite{Amelino-Camelia:2016ohi},
 \cite{Xu:2018ien}, \cite{Ellis:2018lca} (see also \cite{Acciari:2020kpi} for slightly more stringent lower limits).

Lorentz invariance violating (LIV) theories generically provide
modified dispersion relations; among them there is an interesting class where the Lorentz
group (or its realization) is modified, so that a new relativistic symmetry replaces the
classical one, these theories go under the name of Deformed (or Doubly) special relativity theories
(DSR theories) \cite{AmelinoCamelia:2000mn},
\cite{AmelinoCamelia:2000ge}, \cite{KowalskiGlikman:2001gp}, \cite{MS1}, \cite{MS2}, \cite{JMACP} and are more appealing since they preserve a
relativity principle; moreover, the deformed Lorentz symmetry allows for
milder deviations from special relativity kinematics.
Many of these phenomenological models describe spacetime features and wave equations that are typical of noncommutative spacetimes, the prototypical example being $\kappa$-Minkowski spacetime, where coordinates satisfy the relations
$x_0 \star x_j - x_j \star x_0 = i \kappa\/ x_j , x_i \star x_j - x_j \star x_i = 0$;
here we consider $ 1/\kappa$ to be related to the
Planck length. See \cite{KowalskiGlikman:2001gp}
for an early relation between DSR and $\kappa$-deformed symmetries.
Noncommutative geometry, as the generalised notion of
spacetime geometry were a minimum length emerges due to spacetime
noncommutativity, and were groups of symmetry transformations  are
deformed in  quantum symmetry groups can indeed be helpful in providing
models quantifying the effects of quantum gravity without full knowledge of
quantum gravity itself, but incorporating at a kinematical level the
key dynamical aspect of existence of spacetime uncertainty
relations. Notice that these latter are generically inferred from gedanken experiments probing
spacetime structure at Planck scale and independently arise in String
Theory and as minimal area or volume elements in Loop Quantum Gravity, see e.g. \cite{Hossenfelder:2012jw}.

The interplay of the Planck scale effects on the dispersion relations
in  quantum (noncommutative) spacetime were investigated in the curved backgrounds of expanding
universe in
\cite{constraints}, \cite{Piran}, \cite{RACMM}, \cite{Barcaroli:2016yrl}.

\bigskip 

In this paper we use a top-down approach that complements the
bottom-up one of phenomenological models. We apply noncommutative differential geometry to derive
the propagation of waves in noncommutative cosmology. We study a noncommutative deformation of the wave equation in curved background and we discuss
the modification of dispersion relations due to the presence of both 
noncommutativity and curvature of spacetime. As a
first approximation we turn on noncommutativity in the usual
(classical) homogeneous and isotropic gravity solution given by FLRW
spacetime, and  derive the wave equation for massless particles in this
context. This is a first step toward a more comprehensive approach
that encompasses both the dynamics of light and of gravity in a
noncommutative spacetime. We here consider a classical gravity background.

In \cite{ABP} we have obtained the wave equation on a wide class of
noncommutative spaces deriving it from  first principles associated with noncommutative differential
geometry and the corresponding geometric and physical notion of
noncommutative infinitesimal translations,  i.e. quantum momenta. We have found that contrary
to the generic LIV theories expectations,  in flat spacetime no dispersion
relations arise (but modified Einstein-Planck relations do arise).
The study of dispersion relations in flat spacetime is a first propaedeutical
step in the study of dispersion relations in cosmological
spacetime.  
The present paper can be considered as a sequel to \cite{ABP}, while
there wave equations for massless fields in flat noncommutative spacetime  were considered in the context of
a correspondingly deformed (quantum)
Poincar\'e-Weyl symmetry, here we turn on a
nontrivial curvature and focus our attention to the metric of FLRW cosmology.  
We find that in this curved case modified dispersion relations for
massless fields do indeed arise, the modification is proportional to
$E_{ph}/E_{LV}$ (the travelling photon energy over the Lorentz violation scale
related to Planck energy) to the
cosmological time, to the Hubble parameter $H$ and to the inverse scale
factor $a(t)$. In particular we immediately recover the result of
\cite{ABP}: for flat spacetime,  $H=0$, 
there is no modified dispersion.
\\

We follow the Drinfeld twist formalism that leads to noncommutative
spaces via a star product deformation of their algebra of
functions. It also canonically gives a noncommutative differential
calculus and wave equations \cite{ABP}.
In Section II we recall the geometric construction of the wave equation in
curved noncommutative spacetime this is given by the twist deformed Laplace-Beltrami operator for arbitrary curved metric in the presence of the noncommutativity.  
The deformed wave equation proposed here can be constructed for any
twist and  for any curved background.

In Section III we consider the specific Drinfed twist  called
Jordanian twist  which has built in a minimal length, it is spatially isotropic and induces
a noncommutativity that reduces, in the
limit of  flat spacetime to $\kappa$-Minkowski spacetime with its quantum Poincar\'e-Weyl symmetry
for massless particles; moreover, as shown in \cite{ABP}, in this limit it gives the nonlinear realization of the
Lorentz generators of the Doubly special relativity theories
\cite{MS1}, \cite{MS2}.
The Jordanian twist is defined by, cf. \cite{Borowiec:2008uj}:
\begin{equation}
\mathcal{F}=\exp \left( -iD\otimes \sigma \right) \quad ;\quad \sigma =\ln
\left( 1+\frac{1}{\kappa }P_{0}\right) ~, \label{nsymJ}
\end{equation}%
where $D=-ix^\mu\partial_\mu$ is the dilatation generator and
$P_0=-i\partial_0$ is the time translation generator. For other
realizations of Jordanian twist see e.g. \cite{G-Z,Int1,Int2}.

In Section IV we specialise the metric
to the FLRW one, 
we obtain the first order correction in the
noncommutative deformation parameter  $\kappa$ to the velocity
dispersion relations; the result does not depend on the spacetime
dimensions, we first treat the simpler 2 dimensional case and then the
4 dimensional one. The time arrival lag of energetic photons with
respect to low energetic ones is computed. As a first approximation,
comparing with estimates coming from recent analysis of GRB data
\cite{Xu:2018ien}, \cite{Ellis:2018lca}, 
the net result is the constraint $\hbar\kappa\sim$ few $ 10^{18}$ GeV, that is
one  order of magnitude higher than that obtained via usual modified
dispersion relations that depend only on $E_{ph}/E_{LV}$, and very
close to Planck energy ($1.22\times 10^{19}$ GeV). We then further comment on the general
mechanisms leading to modified dispersion relations when
both noncommutativity and curvature are turned on.
After the conclusions, presented in Section V, we provide
two appendices offering more informations
on the noncommmutative framework considered and the curved spacetime group velocity computation. 

\section{Differential Geometry on $\kappa$-spacetime from twist deformation}

\subsection{$\kappa$-spacetime and Jordanian Twist}
Noncommutative $\kappa-$Minkowski spacetime is the algebra generated by the
coordinates $x^\mu$ ($\mu=0,\ldots n-1$) that satisfy the commutation relations 
\begin{equation}
 x^{0}\st x^{\/j}-x^{\/j}\st x^0=\frac{i}{\kappa }x^{\/j}~,
 \quad \quad x^{i}\st x^{j}-x^j\st x^i=0~,
\label{kmink}
\end{equation}
where $i$ and $j$ run over the space indices $1,\ldots n-1$, while
$\kappa$ is the noncommutativity  parameter and  we have denoted by $\st$
the corresponding noncommutative product.  In this paper we do not fix the metric to be $\eta=diag(-1,1,\ldots 1)$ and hence prefer to
simply refer to \eqref{kmink} as to the defining relations of  $\kappa$-spacetime.

A far reaching way to obtain the commutation relations \eqref{kmink}
is via a Drinfeld twist. In this paper we focus on a specific case
which is called  Jordanian twist $\mathcal{F}$, given in \eqref{nsymJ}. The inverse of this twist is $\mathcal{F}^{-1}=\exp \left( iD\otimes \sigma \right)
=\exp \left( x^\mu\partial_\mu\otimes
  \ln(1-\frac{i}{\kappa}\partial_0) \right) $. 
Due to the algebraic properties of $\mathcal{F}$, that follow from the
Lie algebra $[D, P_0]=iP_0$ of the time translation $P_0$
and of the dilatation $D$, given smooth functions $f$,
$h\in A=C^\infty(\mathbb{R}^n)$, the $\star$-product defined by
\eq\label{starprodfunc}
f\st h=\mu \{\mathcal{F}^{-1}(f\otimes h)\}
\en
i.e., $(f\st h)(x)= \exp \big( x^\mu \frac{\partial}{\partial x^\mu}\otimes
  \ln(1-\frac{i}{\kappa}\frac{\partial}{\partial y^0})
  \big)f(x)h(y)\big|^{}_{x=y}$, is associative. In  \eqref{starprodfunc} $\mu$ is the usual
pointwise product $\mu (f\otimes h)(x)=f(x)h(x)$. We denote by $A_\st$
the algebra of smooth functions on $\mathbb{R}^n$ where the product is
given by the
$\star$-product in \eqref{starprodfunc}.
One  can expand $\mathcal{F}^{-1} $ in power series of
$\frac{1}{\kappa}$, see
\cite{Borowiec:2008uj, ABP}, 
\eq
\mathcal{F}^{-1}=
1\otimes 1 + iD\otimes \frac{1}{\kappa}P_0 +
\frac{1}{2}iD(iD-1)\otimes \frac{1}{k^2}P_0^2\, +\ldots\;
=\,\sum_{n=0}^\infty
\frac{(iD)^{\underline{n}}}{n!}\otimes \big(\frac{1}{\kappa}P_0\big)^n
\label{nsymJ2}\en
where $X^{\underline{n}}=X(X-1)(X-2)\ldots (X-(n-1))$ is the so-called
lower factorial. 
In particular,
we easily see that when 
$f$ and $h$ are coordinate functions,  the commutation relations
\eqref{kmink} hold.

The realization of  $\kappa$-spacetime via a $\star$-product obtained from
a twist ${\cal F}$ allows to readily construct a corresponding
noncommutative differential geometry.

\subsection{Differential calculus}\label{DiffCalc}
It is useful to introduce the following shortcut notation for the twist: $$\mathcal{F}^{-1}=\of^\al\otimes\of_\al~,$$
where sum over the index $\alpha$ is understood (it corresponds 
to the sum in \eqref{nsymJ2}). 
In this notation the deformation of  the algebra $A$ into the algebra
$A_\st$ of $\kappa$-spacetime is given by, cf. \eqref{starprodfunc},
\begin{equation}\label{fsth}
f\st h=\of^\al(f)\of_\al(h)~. 
\end{equation}
Following \cite{GR2} (see also \cite[\S 7.7]{LNP-book} and \cite{ABP}) there is a canonical
construction in order to obtain the algebra of forms and the
exterior differential.
 Similarly to \eqref{fsth},  the algebra of exterior forms
$\Omega^\bullet=A\oplus\Omega^1\oplus\Omega^2\oplus...$ can be
deformed to the algebra $\Omega^\bullet_\st$, which as a vector space
is the same as the undeformed  $\Omega^\bullet$ 
but has the new wedge $\st$-product 
\eq
\omega\wedge_\st\omega'=\of^\al(\omega)\wedge\of_\al(\omega')~;
\en
here the action of $\FF^{-1}=\of^\al\otimes \of_\al$ on forms is via the Lie
derivative $\mathcal{L}$ along the vector fields $D$ and $P_0$ defining
$\FF^{-1}$. Explicitly, $D(\omega)=\mathcal{L}_D\omega$, $D^2(\omega)=\mathcal{L}_D\mathcal{L}_D\omega$,
and iteratively $D^p(\omega)=\mathcal{L}_D D^{p-1}\omega$, and similarly for $P_0$ instead of $D$.
In particular, when $\omega'$ is a zero form $f$, then the wedge product
is usually omitted and correspondingly the wedge $\st$-product
reads $\omega\st f=\of^\al(\omega)\of_\al(f)$.

Since the Lie derivative commutes with the exterior derivative the usual
(undeformed) exterior derivative satisfies the Leibniz rule
$
\mathrm{d}(f\star h) =\mathrm{d}f\star h+f\star \mathrm{d}h,  
$
and more in general, for forms of homogeneous degree $\omega\in \Omega^r$, 
\eq
\d(\omega\wedge_\st\omega')=\d\omega\wedge_\st\omega'+(-1)^r\omega\wedge_\st
\d\omega'~.
\en
We have constructed a differential calculus on the deformed algebra of
exterior forms $\Omega_\st^\bullet$.

For later purposes we compute the differential of a function $f$ as 
\eq\label{dfdxdf}
\mathrm{d}f=\mathrm{d}x^\mu_{~\:} \partial_\mu f=\mathrm{d}x^\mu \star \partial_\mu^\FF f~,
\en 
where in the last expression we have introduced the $\star$-product
between one-forms and functions, and deformed the partial derivative
$\partial_\mu$ into the quantum one defined by
\eq\label{delF}
\partial_\mu^\FF
f=\frac{1}{1-\frac{i}{\kappa}\partial_0}\partial_\mu~.
\en
The proof of \eqref{dfdxdf} easily follows recalling the explicit
expression of the inverse twist.

\section{Wave equations}
In order to formulate dynamical theories we need a metric on
spacetime, equivalently, using the language of exterior forms, we need a
$*$-Hodge operator. We first see how to canonically define the latter in the
noncommmutative setting. Then we present the wave equation on
$\kappa$-noncommutative spacetime with an arbitrary metric.

\subsection{Metric and Hodge star operator}
For an $n-$dimensional manifold with metric $g$ 
the Hodge $\ast $-operation is a linear
map on the space of exterior forms $\ast :\Omega ^{r} \rightarrow \Omega ^{n-r}$. In local coordinates an $r$-form is given by\\
$\omega=\frac{1}{r!}\omega _{\mu _{1....}\mu_{r}}\d x^{\mu_1}\wedge\ldots
\d x^{\mu_r}$  and the Hodge $\ast$-operator reads
\eq\ast \omega =\frac{\sqrt{g}}{r!\left( n-r\right) !}\omega _{\mu _{1....}\mu
_{r}}\epsilon _{~\ \ \ \ \ \ \ \ \ \ \!\!\!\!\!\!\!\!\!\nu _{r+1}......\nu _{n}}^{\mu
_{1....}\mu _{r\,}}\d x^{\nu _{r+1}}\wedge \ldots \d x^{\nu _{n}}~
\en
where $\sqrt{g}$ is the square root of the absolute value of the determinant of the metric, the completely antisymmetric tensor $\epsilon_{\nu_1\ldots
  \nu_n}$ is normalized to $\epsilon_{1\ldots n}=1$ and indices are
lowered and raised with the metric $g$ and its inverse.
There is a one to one correspondence between metrics and Hodge star
operators (indeed $\d x^\mu\wedge * \:\!\d x^\nu=g^{\mu\nu}\,*\!1$).

We define metrics on noncommutative spaces by defining the
corresponding Hodge star operators on the $\st$-algebra of exterior forms
$\Omega_\st^\bullet$. We first observe that the undeformed Hodge
$\ast$-operator is $A$-linear:
$\ast(\omega f)=\ast(\omega)\,\! f$, for any form $\omega$ and function
$f$
(of course, since $A$ is commutative we equivalently  have
$\ast(f\omega)=f\,\!(\ast\omega)$). 
We then require the Hodge $\ast$-operator $\ast^\FF$ on $\Omega_\st^\bullet$ to
map $r$-forms into $(n-r)$-forms, and to be right $A_\st$-linear
\eq\label{eqwf}
\ast^\FF(\omega\star f)=\ast^\FF(\omega)\star f
\en
for any form $\omega$ and function $f$.
The quantum Hodge operator $\ast^\FF$ is then the deformation 
of the usual Hodge $\ast$-operator given by:
 \begin{eqnarray}\label{quantum_ast}
\ast ^{\mathcal{F}}
: \,\Omega_\st^\bullet&\longrightarrow& \Omega^\bullet_\st\nonumber\\
\omega & \longmapsto & 
\ast^{\mathcal{F}}(\omega)
=\mathrm{\bar{f}}^\al(\ast)_{\,}\of_\al(\omega)~.
\end{eqnarray}
In \eqref{quantum_ast} the action 
of $_{}\of^\al$ on the usual Hodge $\ast$-operator is the adjoint
action. Recall that $\of^\al$ for each index $\alpha$ is a polynomial in the dilatation
$D=-ix^\mu\partial_\mu$; then the action of $D$ on the Hodge
$\ast$-operator is defined by $D(\ast)(\omega)=D(\ast\omega)-\ast (D(\omega))$, i.e., $D(\ast)=D\circ
\ast-\ast\circ D=[D,\ast]$. The action of $D^2$ is hence $[D,[D,\ast]]$,
and iteratively $D^p(\ast)=[D,D^{p-1}(\ast)]$.
This defines $\of^\al(\ast)$ for any index $\alpha$. 
From definition \eqref{quantum_ast} it immediately follows that in the
commutative limit $\kappa\to \infty$ we have $\ast^\FF\to \ast$. From the general theory of twist deformation of maps,
cf. \cite{ABP}, \cite[Theorem 4.7]{AS},  it also follows
that for any exterior form $\omega$ and function $f$ we have the right
$A_\st$-linearity property $\ast ^{\mathcal{F}}(\omega\st f)=\ast
^{\mathcal{F}}(\omega)\st f$. We also notice that
definition \eqref{quantum_ast} of quantum Hodge operator parallels that used to define quantum vector fields and the physical quantum momenta $P^
\FF_\mu$, as we review in  Appendix \ref{vectorfields}. 

Finally, we remark that there is no a priori relation between the
metric structure $g$ we have introduced via the Hodge star operator
$\ast$ and the twist $\FF$ determining the noncommutativity of
spacetime. 
We comment more on this at the end of Sec. \S \ref{DNCC}.

\subsection{Wave equations in curved $\kappa$-spacetime}
The wave equation in curved
spacetime is governed by the Laplace-Beltrami operator $$\Box = \delta
d + d\delta~ .$$
In the case of even dimensional Lorentzian
manifolds the adjoint of the exterior derivative is
defined by $\delta= \ast \!\:\mathrm{d}\:\! \ast$.
In particular, for a scalar field we have
\begin{equation}
\Box \varphi =\ast_{\,} \mathrm{d}\!\:\!\ast\!\:\! \mathrm{d}_{\,}\varphi=\frac{1}{\sqrt{g}}%
\partial _{\nu }\left( \sqrt{g}g^{\nu \mu }\partial _{\mu }\varphi \right)~.
\label{thm}
\end{equation}
This is the wave equation for a scalar field minimally coupled to a
background gravitational field.

Wave equations
in noncommutative spacetime are defined by
just replacing the Hodge $\ast$-operator with the
$\ast^\FF$-operator introduced in (\ref{quantum_ast}).  For even dimensional noncommutative spaces with Lorentzian metric: $$\square^\FF=\ast ^{\mathcal{F}}\mathrm{d}\ast ^{\mathcal{F
}}\mathrm{d}+ \mathrm{d}\ast ^{\mathcal{F
}}\mathrm{d}\,\ast ^{\mathcal{F}},$$ hence for a scalar field  we have
\begin{eqnarray} 
\label{scalar_eq}
  &&~~~\square ^{\mathcal{F}}\varphi =\ast ^{\mathcal{F}}\mathrm{d}\ast ^{\mathcal{F
}}\mathrm{d}\varphi =0~.
\end{eqnarray}
Notice that this wave equation can be constructed for any twist and
any curved background. 
Using the specific Jordanian twist we can explicitly compute
\eqref{scalar_eq}.  From the definition of Hodge $\ast$-operator we have
\begin{eqnarray}\label{astdxdxdx} \ast^\FF(\d x^1\wedge_\st....\d x^r)=\ast(\d x^1\wedge_\st....\d
x^r)~,
\end{eqnarray}
 i.e., on these forms it equals the commutative Hodge
$\ast$-operator associated with an arbitrary curved metric. Indeed,
since for the Jordanian twist each term $\of_\al$ in the second leg of
the tensor product 
$\FF^{-1} =\of^\al\otimes\of_\al$
is a power of $P_0$ and
$P_0 (\d x^\mu)=0$,  it is immediate to see that
$$\d x^1\wedge_\st....\d x^r=\d x^1\wedge....\d x^r$$ and  
$P_0(\d x^1\wedge_\st....\d x^r)= 0,$
hence
$\of^\al\otimes \of_\al (\d x^1\wedge_\st....\d x^r)=1\otimes  \d
x^1\wedge_\st....\d x^r$, and therefore 
$$\ast^\FF(\d x^1\wedge_\st....\d x^r)=\of^\al(\ast)\of_\al(\d x^1\wedge_\st....\d 
x^r)=\ast(\d x^1\wedge_\st....\d x^r)~.$$

Because of right $\st$-linearity,
$\ast^\FF(\d x^1\wedge_\st....\d x^r\st f)=\ast(\d x^1\wedge_\st....\d
x^r)\st f$, recalling \eqref{dfdxdf} it follows that
\begin{equation}
\ast^\FF(\d \varphi)=\ast^\FF(\d x^\mu\star \partial_\mu^\FF\varphi)=
\ast^\FF(\d x^\mu)\star \partial_\mu^\FF\varphi=
\ast(\d x^\mu)\st \frac{1}{1-\frac{i}{\kappa}\partial_0}\partial
_{\mu}\varphi~,
\end{equation}
henceforth
\begin{eqnarray}\label{d*Fd}
\d\big( \ast^\FF(\d \varphi)\big)&=&\d \big(\ast(\d x^\mu)\big)\st
                                \frac{1}{1-\frac{i}{\kappa}\partial_0}\partial_{\mu}\varphi
\:+\: (-1)^{n-1}\ast(\d x^\mu)\wedge_\st
                                     \d\big(\frac{1}{1-\frac{i}{\kappa}\partial_0}\partial_{\mu}\varphi\big)\nonumber\\
&=&\big(\frac{1}{(n-1)!} \partial_\rho(\sqrt{g} g^{\mu\nu})\varepsilon_{\nu\nu_1...\nu_{n-1}}\d x^\rho\wedge
    \d x^{\nu_1}...\wedge\d x^{\nu_{n-1}}\big)\st \frac{1}{1-\frac{i}{\kappa}\partial_0}\partial
_{\mu}\varphi\nonumber\\
& &~ +\, (-1)^{n-1}\big(\frac{1}{(n-1)!} \sqrt{g} g^{\mu\nu}\varepsilon_{\nu\nu_1...\nu_{n-1}}
    \d x^{\nu_1}...\wedge\d x^{\nu_{n-1}}\big)\wedge_\st \frac{1}{1-\frac{i}{\kappa}\partial_0}\partial_\rho\partial_{\mu}\varphi\,\d x^\rho\nonumber\\
&=&\Big(\partial_\nu (\sqrt{g} g^{\mu\nu})\st
    (1-\frac{i}{\kappa}\partial_0)^{n-1}\partial_\mu\varphi+
\sqrt{g} g^{\mu\nu}\st   (1-\frac{i}{\kappa}\partial_0)^{n-2}\partial_\nu\partial_\mu\varphi\Big)\st
(\d x^1\wedge\d x^2... \d x^n).\nonumber\\
\mbox{}\end{eqnarray}
Here in the last passage we have used that $$\sqrt{g} g^{\mu\nu}\varepsilon_{\nu\nu_1...\nu_{n-1}}\d x^\rho\wedge
    \d x^{\nu_1}...\wedge\d x^{\nu_{n-1}}=\sqrt{g} g^{\mu\nu}\st\varepsilon_{\nu\nu_1...\nu_{n-1}}\d x^\rho\wedge_\st
    \d x^{\nu_1}...\wedge_\st\d x^{\nu_{n-1}},$$ then we have moved $\frac{1}{1-\frac{i}{\kappa}\partial_0}\partial_{\mu}\varphi$ to the
    left,
    and similarly for the other addend. Finally we rewrote $\d x^\rho\wedge
    \d x^{\nu_1}...\wedge\d
    x^{\nu_{n-1}}=\varepsilon_{\rho\nu_1...\nu_{n-1}}\d x^1\wedge\d
    x^2...\wedge \d x^n$ and performed the usual
    epsilon tensor contractions.

From the invertibility of the Hodge $*^\FF$ operator, (or directly
moving in (\ref{d*Fd})
the $n$-form\\ $\d x^{1}\wedge \d x^2 ... \,\d x^{{n}}$  to the left and then
applying the Hodge $*^\FF$ operator) we see that the $n$-dimensional wave equation
$\Box^{\mathcal{F}}\varphi =\ast^{\mathcal{F}} \d\ast^{\mathcal{F}}
\d\varphi =0$ in the presence of $\kappa$-noncommutative spacetime and
with arbitrary curved metric is equivalent to $\d\ast^{\mathcal{F}}\d\varphi =0$ and to
\begin{equation}\label{LBeq*}
\sqrt{g} g^{\mu\nu}\st 
(1-\frac{i}{\kappa}\partial_0)^{n-2}\partial_\nu\partial_\mu\varphi+\partial_\nu (\sqrt{g} g^{\mu\nu})\st 
    (1-\frac{i}{\kappa}\partial_0)^{n-1}\partial_\mu\varphi=0~. 
\end{equation}
\\[-1em]

We conclude this section observing that if we consider the usual Minkowski metric
$g^{\mu\nu}=\eta^{\mu\nu}=diag(-1,1\ldots 1)$ 
then
\eqref{LBeq*} gives the wave equation in $\kappa$-Minkowski
spacetime studied in  \cite{ABP}. This wave equation is equivalent to the wave equation
constructed from the quadratic quantum Casimir operator $ {P^\mu}^\FF
P_\mu^\FF $, see Appendix \ref{vectorfields} for the definition of the
quantum momenta $P_\mu^\FF$; it is also equivalent to the wave
equation constructed from the canonical twist deformation $\square\mapsto
\square^\FF={\cal D}(\square)=\of^\al(\square)\of_\al$ of the d'Alembert
operator of usual Minkowski spacetime, see \cite{ABP}. 
For massless fields this wave equation is also
equivalent to the usual wave equation in
commutative Minkowski spacetime; its physical interpretation
leads to unmodified dispersion relations and to modified Einstein-Planck
relation between energy and frequency \cite{ABP}.
\sk

\section{Dispersion relations in $\kappa$-noncommutative cosmology} \label{DNCC}

We study a theoretical model based on first principles leading to
massless fields dispersion relations and focus on the case of  gamma ray bursts and
the time delay between high energy and low energy photons.
The natural setting is that of a distant source that
emits a gamma ray burst, emitter and observer in first
approximation do not have peculiar velocities and can be considered at rest
with respect to the usual comoving coordinate system $(t,x^i)$ of FLRW
cosmology, where
$\mathrm{d}s^{2}=-\mathrm{d}t^{2}+a(t)^{2}\mathrm{d}x^{2}$.

In this section we study a model in 2 and in  4 dimensions and guided
by the results obtained we extrapolate general considerations.
We show that when the $(t,x^i)$ coordinates become noncommutative the speed of
propagation of a massless scalar field --i.e., neglecting spin, the
speed of light-- depends on the energy, on the cosmic time and on the expansion
rate; hence nontrivial dispersion relations occur.
These are due to the interaction between spacetime curvature and its
noncommutativity, indeed, if the background is curved and commutative, or if it
is flat and noncommutative, as we have shown in \cite{ABP}, there is no dispersion relation. %
\vskip.4cm

We implement noncommutativity of FLRW
spacetime $\mathrm{d}s^{2}=-\mathrm{d}t^{2}+a(t)^{2}\mathrm{d}x^{2}$
by prescribing the commutation relations of the comoving coordinates
$(t,x^i)$. Let's recall that $t$ is the time measured by a clock in
position $x^i$ (a comoving observer in $x^i$), so that $\mathrm{d}t$ captures a
local property of spacetime in region $(t,x^i)$, while $x^i$ is the
rescaled  distance so that velocity of light is
$\mathrm{d}|x|/\mathrm{d}t=\frac{1}{a(t)}$.
Changing perspective between metric and noncommutative structures, we
can say that we implement curvature in
$\kappa$-spacetime by identifying the coordinates of
$\kappa$-spacetime as comoving coordinates.
 \\

\subsection{Scalar field in 2 dimensions}
We 
first specialize (for simplicity) to the 2 dimensional FLRW spacetime with metric
$g^{\mu\nu}=(-1, 1/a(t)^2)$,  the wave equation in \eqref{LBeq*} then reads
\begin{equation}\label{2dscalar}
a\st\partial_0^{2}\varphi -a^{-1}\st\partial_x^2\varphi +
(\partial_0 a)\st
    (1-\frac{i}{\kappa}\partial_0)\partial_0\varphi =0~.
\end{equation}
In order to solve this equation and study the corresponding dispersion
relations we proceed in analogy with the commutative case, that is
propaedeutical and reviewed in
Appendix \ref{GVNC}. We set $\varphi_k=\lambda(t)\st
e^{-ikx}=\lambda(t) e^{-ikx}$ so that $\partial^2_x\varphi=-k^2\varphi$
and the equation simplifies to:
\begin{equation}
a\st \partial_0^2\la+ (\partial _{0}a) \star  (1-\frac{i}{\kappa}\partial_0)\partial
_{0}\lambda  + k^{2}  a^{-1}\star\lambda =0~.
\label{eq1simplified}
\end{equation}
We study this equation at the first order in the noncommutativity
parameter $\frac{1}{%
\kappa }$; its expansion, using \eqref{nsymJ2}, explicitly reads:
\begin{equation}
a \,\partial _{0}^{2}\lambda +\partial _{0}a \big( 1-\frac{i}{%
\kappa }\partial _{0}\big) \partial _{0}\lambda +
k^{2}a^{-1}\lambda -\frac{i}{\kappa }t\left( \partial _{0}a \,\partial
_{0}^{3}\lambda +\partial _{0}^{2}a \,\partial _{0}^{2}\lambda + k^{2}\partial
_{0}a^{-1}\partial _{0}\lambda \right) =0~.
\end{equation}%

Since this equation is a deformation of the wave equation in
commutative FLRW spacetime, following that case we change the
time coordinate into conformal time $\eta$.  We use the relations $\partial
_{0}=\frac{1}{a}\partial _{\eta };\partial _{0}^{2}=-\frac{a^{\prime }}{a^{3}%
}\partial _{\eta }+\frac{1}{a^{2}}\partial _{\eta }^{2};\partial _{0}^{3}=%
\frac{1}{a}\left( \frac{3\left( a^{\prime }\right) ^{2}}{a^{4}}-\frac{%
a^{\prime \prime }}{a^{3}}\right) \partial _{\eta }-\frac{3a^{\prime }}{a^{4}%
}\partial _{\eta }^{2}+\frac{1}{a^{3}}\partial _{\eta }^{3}$
and introduce the simplifying notation $s=\ln a;s^{\prime }=\frac{a^{\prime }}{a};%
\frac{a^{\prime \prime }}{a}=s^{\prime \prime }+\left( s^{\prime }\right)
^{2}$, where the prime denotes the derivative $\partial/\partial\eta$. This results in:
\begin{equation}\label{eqforla}
\frac{1}{a}\lambda ^{\prime \prime }+\frac{k^{2}}{a}\lambda -\frac{i}{\kappa
a^{3}}t\left( \eta \right) \left( \big(2 (s')^{3}-2s^{\prime }s^{\prime \prime }-k^{2} s^{\prime }\right) \lambda ^{\prime}+\left( s^{\prime \prime }-3\left( s^{\prime }\right) ^{2}\right) \lambda
^{\prime \prime }+s^{\prime }\lambda^{\prime \prime \prime }) -\frac{i}{\kappa a^{2}}s^{\prime }\left( -s^{\prime }\lambda ^{\prime }+\lambda
^{\prime \prime }\right) =0~.
\end{equation}%
We substitute
\begin{equation}\label{laF}
\lambda =\exp \Big( i\omega \eta +\frac{i}{\kappa }F\Big)
\end{equation}
and at zero-th order in the
deformation parameter we obtain $\left( \omega
^{2}-k^{2}\right) \lambda =0$, that we solve choosing $\omega=k$
(corresponding to a forward travelling wave),
while at the first order in $\frac{1}{\kappa}$, using the zero-th order
solution $\omega=k$,
we obtain the  differential equation for $F(\eta) $:
\begin{equation}
F^{\prime \prime }+2i k F^{\prime }=\frac{ik t(\eta) }{%
a^{2}}\left( 2\left( s^{\prime }\right) ^{3}-2s^{\prime }s^{\prime \prime
}-2 k^{2} s^{\prime }+i k \big( s^{\prime \prime }-3\left( s^{\prime
}\right) ^{2}\big)\right) -\frac{ik}{a}%
s^{\prime }\left( s^{\prime }-ik \right)  \label{diff_eq_F}~.
\end{equation}%
We aim at the expression of the group velocity for the wave
\begin{equation}\label{gvnc}
\varphi_k(x,t)=\la(t)\star e^{-ikx}=\la(t)e^{-ikx}=\exp \Big( i k \eta +\frac{i}{\kappa }F\Big) e^{-ikx} =e^{i(f_k(t)-kx)}
\end{equation}
where the last equality defines $f_{k}\left( t\right) =\left( k \eta
  +\frac{1}{\kappa }F\right)\!(t) $. Recalling the group velocity
expression 
$
v_{g}=\frac{\partial x}{\partial t}=\frac{\partial }{\partial k}\frac{\partial f_k(t)}{\partial t}
$, cf. (\ref{gv}) in Appendix \ref{GVNC}, we see that we need to
compute $\dot{F}=\partial F/\partial t$.

This is easily obtained from the differential equation
(\ref{diff_eq_F}) in the physical regime we are interested in: cosmic
time related to large scale structure formation, and high frequency
waves. There are three frequency parameters in the differential
equation (\ref{diff_eq_F}): $\omega=k$, $t^{-1}$ and the Hubble
parameter $H$; we obviously have  $\omega >>t^{-1}$ for the present
cosmic time as well as the cosmic time of emission of the travelling
$\gamma$-ray, typically at redshift $z=a^{-1}-1$ below $z=10$.
Similarly  $\omega>>H\sim t^{-1}$.

In this regime (\ref{diff_eq_F}) 
   simplifies to	$2ikF'=-\frac{2ik^3ts'}{a^2}$,
   i.e.\footnote{An easy consistency check is to compute $F''$ from
     \eqref{Fprimo} and explicitly see that it is negligible with
     respect to $kF'$.
It is also instructive to consider, as an example, a single
  fluid evolutionary scenario with the normalized scale factor $a(t)=
  (t/t_0)^\alpha, \alpha =\frac{2}{3(1+w)}$, where $w$ is a barotropic
  factor ($w={1\over 3}, \alpha=1/2$ -- radiation, $w=0, \alpha=2/3$
  -- dust/Dark Matter dominated era) and $t_0$ denotes the age of our
  Universe. 
Due to the normalization  condition the scale factor is of order $1$,
i.e. $a(t) \sim 1$, and $t_0\sim H_0^{-1}\sim 10^{18}$ {s}. Thus $\frac{t}{a^2}\sim
H_0^{-1}$. Similarly, in the conformal time,
$a(\eta)=(\eta/\eta_0)^\beta, \beta=\frac{2}{1+3w}$ and the function $s'$ is of the
order of $H_0$ while $s''$ is  at most
few orders of magnitude bigger than $H_0^2$ (for the low redshifts $z$
of the gamma ray bursts events of interest). Now, because of the
extremely small value of the Hubble constant $H_0$ with respect to the frequency
$\omega=k$, the only term which survives  this approximation on the r.h.s. of (\ref{diff_eq_F}) is $\frac{k^3 t s'}{a^2}\sim k^2$, while, on the l.h.s., it is $2ikF'$.}
\eq \label{Fprimo}
\dot{F}=-\frac{k^2t\dot{a}}{a^3}~.
\en
Hence the group velocity,
at the first order in the $\frac{1}{\kappa}$ deformation,  results
\begin{equation}\label{groupdisprel}
v_g=\frac{\partial x}{\partial t}=
\frac{\partial }{\partial k}\frac{\partial f_k(t)}{\partial
  t}=\frac{1}{a}+\frac{1}{\kappa}\frac{\partial \dot{F}}{\partial
  k}=\frac{1}{a}\Big(1-\frac{2}{\kappa}\frac{kt\dot{a}}{a^2}\Big)=
\frac{1}{a}\Big(1-\frac{2}{\kappa}\frac{\omega t\dot{a}}{a^2}\Big)~.
\end{equation}
In the last equality we have expressed the group velocity in terms of
the frequency $\omega$; this last expression is easily seen to hold also if at zero-th
order in $\frac{1}{\kappa}$ we consider the backward travelling
wave solution $\omega=-k$.  Taking into account the $\frac{1}{a}$ factor due to the comoving
coordinates and inserting the flat spacetime speed of light $c$ we see
that $\kappa$-spacetime  
noncommutativity in the presence of a FLRW metric leads to a physical velocity
of massless scalar 2d particles $v_{ph}=v_ga$ given
by  
\begin{equation}\label{vph2d}
v_{ph}=c(1-\frac{2}{\kappa}\frac{\omega t\dot{a}}{a^2})~.
\end{equation}

\subsection{Scalar field in 4 dimensions}
In the 2 dimensional commutative case the scalar field equation $\Box
\varphi =\ast_{\,} \mathrm{d}\!\:\!\ast\!\:\!
\mathrm{d}_{\,}\varphi=0$ describes a minimal coupling to the
curvature (there is no term $R\varphi$ proportional to the scalar
curvature $R$); this is also a  conformal coupling (if $\varphi$ is a
solution with metric $g_{\mu\nu}$ it is also a solution with conformally
rescaled metric $\Omega(t,x)g_{\mu\nu}$). 
In the 4d case, as in the 2d case, and since electromagnetism couples conformally to the
metric,  we consider a scalar field conformally coupled to gravity,
hence we add to the 4d noncommutative scalar field equation
\eqref{LBeq*} a term that is proportional to the commutative scalar curvature, 
\begin{equation}\label{LBeqR*}
\sqrt{g} g^{\mu\nu}\st 
(1-\frac{i}{\kappa}\partial_0)^{2}\partial_\nu\partial_\mu\varphi+\partial_\nu (\sqrt{g} g^{\mu\nu})\st 
    (1-\frac{i}{\kappa}\partial_0)^{3}\partial_\mu\varphi-\frac{1}{6}(\sqrt{g}R)\star\varphi=0~. 
\end{equation}
A rigorous derivation of the last addend requires a study of
noncommutative gravity coupled to noncommutative scalar fields. The
expression chosen reduces to the usual conformal coupling in the
commutative limit (cf. e.g. \cite{Jacobson}) and allows to write the wave equation as an
operator $\star$-acting on $\varphi$. Other possibilities, like
$-\frac{1}{6}\sqrt{g}\star R\star\varphi$, can be considered, but
we will see that they
do not affect the scalar field dispersion relations in the regime we
are interested in.

Similarly to the 2d case, we set $\varphi=\lambda e^{-i(k_xx+k_y
  y+k_zz)}=\lambda\star e^{-i(k_xx+k_y y+k_zz)}$, insert
the values of $g^{\mu\nu}={\rm{diag}}(-1,a^2,a^2,a^2)$ and
$\sqrt{g}=a^3$ in \eqref{LBeqR*} and
obtain 
\begin{equation}
a^3\st 
(1-\frac{i}{\kappa}\partial_0)^{2}\partial_0^2\lambda+
k^2 a \star (1-\frac{i}{\kappa}\partial_0)^{2}\lambda+
(\partial_0 a^3)\st 
    (1-\frac{i}{\kappa}\partial_0)^3\partial_0\lambda+\frac{1}{6}(a^3R)\star \lambda=0~,
\end{equation}
  where
  $k=\sqrt{k_x^2+k_y^2+k_z^2}$.
We next expand the powers of  $(1-\frac{i}{\kappa}\partial_0)$ and the
$\star$-product at leading order in
$\frac{1}{\kappa}$,
\begin{eqnarray}\label{4dscalar}
a^3 \partial_0^2\lambda+
k^2 a \lambda+
(\partial_0a^3) \partial_0\lambda+\frac{1}{6}a^3R\lambda
  -\frac{i}{\kappa}(2a^3\partial_0^3\lambda+2k^2 a\partial_0 \lambda+
{3}(\partial_0a^3) \partial^2_0\lambda)+~~~~~~~~
\nonumber\\-\frac{it}{\kappa}\big((\partial_0a^3)\partial_0^3\lambda+k^2(\partial_0a)\partial_0\lambda+
(\partial_0^2
a^3) \partial^2_0\lambda+\frac{1}{6}\partial_0(a^3R)\partial_0\lambda\big)=0~.
\end{eqnarray}
Setting
$$\lambda=a^{-1}e^{i\omega\eta+\frac{i}{\kappa}F(t)}=a^{-1}e^{ik\eta}\big(1+\frac{i}{\kappa}F(t)\big)+{\cal{O}}\big(\frac{1}{\kappa^2}\big)~,$$
where $\eta$ is the conformal time coordinate,
we observe that, since $a^{-1}e^{i\omega\eta}$ with $\omega=k$ is the classical solution to
the 4d scalar field  conformally coupled to FLRW cosmology  (cf.  Appendix \ref{GVNC}), the sum of
the first four addends vanishes as long as $\partial_0=\frac{1}{a}\partial_\eta$ does
not hit $F(t)$. We are then left with terms at leading order in the
noncommutativity deformation $\frac{1}{\kappa}$ and here, as in the 2d
case, using the physical regime $\omega=k>>H\sim t^{-1}$, 
we consider only the leading order in $k$, given by $\partial_0$
hitting $e^{ik\eta}$, for example the first addend just gives $\frac{i}{\kappa}2a^3\partial_0(a^{-1}e^{ik\eta})\partial_0{F}$. Equation \eqref{4dscalar} thus becomes
\begin{eqnarray*}
2a^3\partial_0(a^{-1}e^{ik\eta})\partial_0{F}-2a^3(\frac{1}{a}\partial_\eta)^3(a^{-1}e^{ik\eta})-2k^2\partial_\eta(a^{-1}e^{ik\eta})+~~~~~~~~~~~~~~~~~~~~~~~~~~~~~~~~\\-t\big((\partial_0a^3)(\frac{1}{a}\partial_\eta)^3(a^{-1}e^{ik\eta})+k^2(\partial_0 a)\frac{1}{a}\partial_\eta(a^{-1}e^{ik\eta})\big)=0~. 
\end{eqnarray*}
Again, the time derivatives give non-negligible terms only if they hit
$e^{ik\eta}$ rather than the conformal factor $a$; it follows that the second and
third addend are proportional to $a^{-1}k^3$ and cancel out, so that
the equation simplifies to
$2aik\dot{F}-t(-2\frac{\dot{a}}{a^2}ik^3)=0$, hence we obtain
\begin{equation}\label{4dF}
\dot{F}=-\frac{k^2 t\dot{a}}{a^3}~.
\end{equation}
This is the same equation as the 2d case one \eqref{Fprimo}.  Hence from
the expression of the group velocity of a wave packet in 4d, $\boldsymbol{v}_g=\frac{\partial \boldsymbol x}{\partial
  t}=\frac{\partial}{\partial t}\nabla(ik\eta+\frac{i}{\kappa} F)$, (cf. Appendix
\ref{GVNC})  the modulus $v_g$ of the
group velocity of the 4d scalar field at first order in the
$\frac{1}{\kappa}$ deformation is given in \eqref{groupdisprel},
\begin{equation}\label{groupdisprel4}
{v_g}=\frac{1}{a}\big(1-\frac{2}{\kappa}\frac{\omega t \dot a}{a^2}\big)~.
\end{equation}
Taking into account the $\frac{1}{a}$ factor due to the comoving
coordinates we arrive at a physical velocity
of massless scalar 4d particles $v_{ph}=v_ga$ given
by  
\begin{equation}\label{vph4d}
v_{ph}=c(1-\frac{2}{\kappa}\frac{\omega t\dot{a}}{a^2})
\end{equation}
as in the 2d case (cf. \eqref{vph2d}). As usual we  define the energy
where classical Lorentz 
violation (or better in our case Lorentz deformation) is manifested 
$E_{LV}:=|\kappa| \hbar$. We extrapolate the 2d and 4d results of
scalar fields conformally coupled to curved spacetime to spin one
fields and hence assume that also photons in $\kappa$-noncommutative
FLRW spacetime have the same dispersion relations  \eqref{vph4d}. The variation of 
the physical speed of light $v_{ph}$ with respect to the usual one $c$ (of 
photons in flat spacetime, or of low energetic photons)  is then given by 
\begin{equation}\label{vphc}
| 1-v_{ph}/c|~\sim ~\frac{E_{ph}}{E_{LV}}\frac{2t\dot{a}}{a^2} ~~ . 
\end{equation}
\\
\indent We finally come to the other scalar curvature couplings in \eqref{LBeqR*}, like 
$-\frac{1}{6}\sqrt{g}\star R\star\varphi$, $-\frac{1}{6}\sqrt{g}
R (1-\frac{i}{\kappa}\partial_0)\st\varphi$, or $-\frac{1}{6}\sqrt{g}
R (1-\frac{i}{\kappa}\partial_0)^{2}\st\varphi$. In all these cases the
first order in $\frac{1}{\kappa}$ of these terms, in the regime 
 $\omega=k>>H\sim t^{-1}$ is always negligible with respect to 
 those proportional to $k\dot{F}$ or  $k^3$, therefore the group
 velocity and dispersion relations results \eqref{groupdisprel4}-\eqref{vphc} are
 independent from the ambiguities in the coupling to the scalar curvature.  

\subsection{Physical considerations}
{}We begin listing a few comments:
\begin{itemize}
\item The combined effects of noncommutativity and gravity
  affect the velocity of light by a term linearly dependent on the
  frequency $\omega$, the cosmic time $t$, the Hubble parameter
  $H=\dot{a}/a$ and inversely proportional to the scale factor. We have $v_{ph}<c$ for $\frac{1}{\kappa}$ a positive time
  (as it is usually considered, and in an expansion phase of the universe
  $\dot{a}>0$). 
 If on the other hand we consider the commutation relations $t\st
 x-x\st t=-|\frac{1}{\kappa}|ix$ then $v_{ph}>c$ and the opposite conclusions hold.
In flat spacetime ($\dot{a}=0$) as well as in commutative spacetime
($\kappa\to \infty$) there are no modified dispersion relations.

\item
This result offers an explicit cosmological correction to the usually
considered models, which assume, as the leading power for the correction to the light speed, the expression
$v_{ph}~\sim ~ c(1 - \frac{E_{ph}}{E_{LV}})$  \cite{ACEMNS}.
 It is actually interesting to estimate the fractional variation \eqref{vphc}
of the speed of light, that in terms of the redshift reads
\begin{equation}\label{vphtH}
\delta v/c\equiv|1-v_{ph}/c|\sim 2 (1+z)tH\,E_{ph}/E_{LV}~.
\end{equation}
For example, the most energetic photons
detected by Fermi-LAT from GRB 080916C have measured energy
$E_{ph}=13.2 ~GeV $  (cf. eg. [8]).
Assuming that $\kappa$ is the Planck
mass, the Lorentz deformation scale corresponds to the Planck energy
scale $E_{LV} = 1.22 \times 10^{19}  ~GeV$ and we obtain the
fractional decrease in velocity $\delta v/c=2.15\times 10^{-18}$
 for these energetic photons at time of detection ($z=0$,
 $tH=t_0H_0=13.29 \ Gyr\times 73\ \frac{km/s}{Mpc}$, according to
 $\Lambda$CDM model). These same
 photons at time of emission, i.e.,  at
 redshift $z=4.35$, had energy $(1+z)\times 13.2 ~GeV $ and 
fractional decrease in velocity $\delta v/c=41.6\times 10^{-18}$.

\item We can also study the time lag $\Delta t$ between the arrival of a low
energetic and a high energetic photon emitted simultaneously during a gamma ray burst.
Following \cite{Piran} we observe that the comoving distance
between the gamma ray burst and the observer is the same for both
photons; for the high energy photon it reads
$\int_{t_{em}}^{t_{0}+\Delta t} v_g\, dt$, where $v_g$ is given by
\eqref{groupdisprel}, while for the low energy one it reduces to $\int_{t_{em}}^{t_{0}}
\frac{c}{a}\, dt$. Equating these distances, and considering only
first order corrections we obtain that the time delay $\Delta t$ is
given by
\begin{equation}
\Delta t=\frac{2E_{ph}}{E_{LV}}\int^{t_0}_{t_{em}} \frac{t
  \dot{a}}{a^3}dt=
\frac{2E_{ph}}{E_{LV}}\int^{z}_{0} t\,(1+z')\/dz'~.
\end{equation}
For the range of redshifts we are interested into (up to $z\sim 10$) we
can use the analytic solution $a(t)=(1+z)^{-1}=(\frac{\Omega_m}{\Omega_\Lambda})^{1/3}
\sinh^{2/3}(t/t_\Lambda)$, 
$t_\Lambda=\frac{2}{3H_0\sqrt{\Omega_\Lambda}}$ and obtain the time lag
\begin{equation}\label{quant2}
\Delta t= 2 \frac{E_{ph}}{E_{LV}}t_\Lambda \int_0^{z}
{\rm arcsinh}\sqrt{\frac{\Omega_\Lambda}{\Omega_m}(1+z')^{-3}\,}\, (1+z')dz'~.
\end{equation}
This differs
from  $\Delta' t= \frac{E_{ph}}{E_{LV}}\frac{1}{H_0}\int_0^z
\frac{ (1+z')dz'}{\sqrt{\Omega_m(1+z)^{3}+\Omega_\Lambda}}$, which is
the typical time lag considered
in the literature for the correction to
the dispersion relations induced by a linear Lorentz invariance
violation \cite{Piran}.
Our model gives a time lag that is  $\sim 3$ times this latter (we use
$\Omega_m=0.27$ for the matter density parameter and
$\Omega_\Lambda=0.73$ for the cosmological constant density).

\item The proposed expression for the group velocity $v_g$
has been derived from a 2 as well as from a 4 dimensional wave equation and well describes the
potentialities of the noncommutative theory developed from first
principles: because of specific quantitative predictions like
(\ref{vphc})-(\ref{quant2}) it can be used to constrain the
noncommutativity parameter $\kappa$ and test the model.
For example the GRB data analyses in \cite{Xu:2018ien}, \cite{Ellis:2018lca}
estimate a Lorentz violation energy at the scales $3.6  \times 10^{17}$ GeV and
$\sim 10^{18}$ GeV respectively.  Taking into account the
factor $\sim 3$ due to \eqref{quant2}, we obtain
$\hbar\kappa\sim$ few $10^{18}$ GeV, i.e., very close to Planck
Energy. \phantom{It would be}
\phantom{the}It would be interesting to further investigate the phenomenological implications of
the present model, and to consider spin one massless fields. Indeed in the
comparison with GRBs data we have extrapolated that the speed of light
is the same as that of massless scalar fields conformally coupled to gravity. This is supported by the independence of our
results from the change of dimension and the fact that the wave
equation for the scalar field in 2d can be seen as that of
the scalar gauge potential for electromagnetism in 2d, indeed in the
commutative case, by defining the 1-form $F=\d\varphi$,  we have that locally $\d\ast \d\varphi=0$ is equivalent
to $\d \ast F=0$ and $\d F=0$.

\item In the present work, as a first approximation, we have
  considered a commutative gravity background, hence noncommutativity
  affects only propagation of light. In a noncommutative
  theory of gravity consistently coupled to light, see e.g. \cite{Castellani-Aschieri},
  one could consider the backreaction effects of turning on
  noncommutativity also on the gravitational field. 
\end{itemize}

The result that the combined effects of noncommutativity and
curvature produce modified dispersion relations is expected to be a general
feature of wave equations in noncommutative curved spacetime. The
interaction between noncommutativity and curvature responsible for the
modified dispersion relations can be traced back to the fact that the
vector fields composing the twist are not Killing vector fields for
the metric. Indeed Killing vector fields trivially act on the metric
and therefore the corresponding twist acts also trivially on the
metric or on the Hodge $\ast$-operator.  This leads to wave equations
that are undeformed. The same conclusion holds more in general if
the twist is composed by affine Killing vector fields, i.e.
vector fields  $X$ such that $X(g)=\lambda g$ with $\lambda$ a constant.
This is so because the equations for  massless particles
are independent from rescalings of the metric.
Explicitly, $\ast^\FF={\cal D}(\ast)=\of^\al(\ast)\of_\al$ is
proportional  to $\ast$ if $\of^\al$ contains only affine Killing vector fields.

 An example of this is
provided by the wave equation in $\kappa$-Minkowski spacetime, indeed
the dilatation $D$ present in the Jordanian twist is an affine Killing
vector field (while the time translation $P_0$ is Killing) for the
usual Minkowski metric, and indeed that wave equation is undeformed \cite{ABP}. These observations
parallel those for the gravity field equations considered in
\cite{GS}. On the basis of these observations, while modified dispersion relations are a generic feature
of curved and noncommutative spacetimes, one can concoct examples of
flat spacetimes with modified dispersion relations (considering vector
fields in the twist that are neither Killing nor affine Killing) as
well as examples of curved and noncommutative spacetimes with
unmodified dispersion relations (provided the curved spacetime metric admits
affine Killing vector fields). 

We do not have arguments to support a direct
correlation between the large scale structure of spacetime, given by
the metric via a cosmological solution to Einstein equations, and the quantum
spacetime structure possibly due to local quantum gravity effects. 
Actually in general relativity different cosmological solutions  are compatible  
with the (classical) local spacetime structure, hence it is natural to assume, as a first approximation, that
the metric structure and the noncommutative one are
uncorrelated, and 
therefore to consider twists with no (affine) Killing vector fields for the metric.
Flat spacetime on the other hand captures local properties of
spacetime structure and quantum gravity effects in this background
might result in a noncommutative spacetime structure where
noncommutativity and the Minkowski metric are compatible. The examples of
$\kappa$-Minkowski spacetime (that has Killing and affine Killing
vector fields) and of $\kappa$-FLRW spacetime (that is without such
vector fields) are
according to these  general expectations.

In conclusion, it seems reasonable to first 
consider noncommutativity of flat Minkowski spacetime, test it
against experimental constraints and then extend the model to curved
noncomutative spacetime assuming nontrivial interaction between
noncommutativity and curvature. 
Among the three main kinds of noncommutativity, defined by the
deformation parameter being massles, of mass dimension one or two: the
canonical noncommutativity $x^\mu\star x^\nu-x^\nu\star x^\mu=i
\theta^{\mu\nu}$, the  Lie algebra-type noncommutativity $x^\mu\star x^\nu-x^\nu\star x^\mu=i
C^{\mu\nu}{}_\rho x^\rho$, and the quadratic one $x^\mu \star x^\nu=\Lambda^{\mu\nu}{}_{\rho\sigma}x^\rho
\star x^\sigma$, the $\kappa$-deformed cosmological spacetime we consider is of the appealing
Lie algebra type that has dimensionful deformation parameter and it is obtained by requiring space commutativity and isotropy.
 The methods developed in this paper however 
are canonically derived from twist differential geometry and
can be applied to other noncommutative and curved spacetime
structures (e.g. with space
anisotropy or with canonical noncommutativity 
$x^\mu\star x^\nu-x^\nu\star x^\mu=i
\theta^{\mu\nu}$)  in order to provide further phenomenological models.
These models would give different dispersion relations depending on
the gravitational background considered. Once more data of time
of flight of GRBs photons becomes available, 
these different models could be tested by comparing the
time lags predictions relative to different GRB sources: close by
versus distant sources, so to test the argument of the integral in the time
lag relation \eqref{quant2} and, in order to test isotropy, sources in different
directions.

\section{Conclusions}

One of the mostly
studied possible phenomenological effects of quantum gravity is the
modifications in wave dispersion. While in \cite{ABP} we have applied the general framework of
noncommutativity arising from twist deformation to the study of
$\kappa$-Minkowski spacetime providing a fresh look on modifications in
dispersion relations, here, with the same canonical
differential geometry methods following from twist deformation, we
have focused on the Friedman-Lemaitre-Robertson-Walker case.
While in flat $\kappa$-Minkowski spacetime  there are no modified
dispersion relations (but there are
modified Einstein-Planck relations) here in curved spacetime we find
modified dispersion relations
and are able to obtain an actual correction to the group
velocity. This is due to the interactions between noncommutativity and
curvature of the spacetime.  Even though the presented result concerns
massless scalar fields  it shows the
potential of this geometrical framework in applications to quantum
gravity phenomenology. 
The natural next step is to investigate the modified dispersion
relations for noncommutative electromagnetism in a FLRW cosmological
background; but equally interesting would be the study of the
dispersion relations in black holes or other curved and noncommutative
backgrounds which can be implemented in the framework here proposed.

\appendix
\section{Quantum vector fields and infinitesimal translations}\label{vectorfields} 
A vector field $u=u^\mu\partial_\mu$ is uniquely defined as a
derivation on the algebra $A=C^\infty(\mathbb{R}^n)$, i.e., a differential operator on
$A$ that satisfies the Leibniz rule.
Similarly, vector fields on the noncommutative algebra $A_\st=C^\infty_\star(\mathbb{R}^n)$ 
are braided derivations (deformed derivations) that satisfy the braided Leibniz rule.
The braiding is related with the so-called universal $R$-matrix
$\mathcal{R}$ with inverse
$\mathcal{R}^{-1}=\mathcal{F}\mathcal{F}_{21}^{-1}$,
where  $\mathcal{F}^{-1}_{21}= \of_\al\otimes\of^\al$ (i.e., we have
flipped the two factors in the tensor product).
The notation for the universal $R$-matrix is analogous to the one for the twist: $\mathcal{R}^{-1}={\bar{R}}^\alpha\otimes{\bar{R}}_\alpha~.$
The algebra $A_\st=C^\infty_\star(\mathbb{R}^n)$ is noncommutative with
the noncommutativity controlled by the $R$-matrix, indeed we have
$$f\star h ={\bar{R}}^\alpha(h)\star {\bar{R}}_\alpha(f)$$
as it is easily seen from  the definition of the $\star$-product in \eqref{starprodfunc}.
We say that the algebra $A_\st$ is braided commutative, the braiding
being provided by the $R$-matrix.

There is a one-to-one correspondence between vector fields on the
commutative algebra $A$ and on the noncommutative algebra $A_\st$.
Any vector field on  $A_\st$ can be written as 
\begin{equation}
u^\FF\!=\DD(u) 
\end{equation}
where $u$ is a vector field on $A$. Here
\eq\label{DonUg}
\DD(u):={\bar{\mathrm{f}}}^\alpha(u){\bar{\mathrm{f}}}_\alpha~,
\en 
and the expression  $\of^\alpha(u)$ denotes the (iterated) Lie
derivative action of the vector
field $D$ entering the twist,
on the vector field $u$. Explicitly, $D(u)=[D,u]$, $D^2(u)=[D,[D,u]]$,
and iteratively $D^p(u)=[D,[D^p(u)]]$.

The vector field $ u^\FF\!=\DD(u) $ satisfies the braided (deformed) Leibniz rule
\eq\label{LeibFF}
u^\FF(f\st h)=u^\FF(f)\st h+\oR^\al(f)\st (\oR_\al(u))^\FF(h)~.
\en
Furthermore, vector fields on $A_\st$ form a braided Lie algebra. The braided
commutator (Lie bracket) is defined by
\begin{equation}  \label{TwistedCom}
[u^\FF,v^\FF]_\mathcal{F}=u^\FF v^\FF-({\bar{R}}^\alpha(v))^\FF_{\,}
({\bar{R}}_\alpha(u))^\FF~
\end{equation}
and it is again a braided vector field. Moreover, it is 
braided-antisymmetric and satisfies the braided-Jacobi identity 
\begin{eqnarray}\label{braid1}
&&[u^\FF,v^\FF]_\mathcal{F} =-[({\bar{R}}^\alpha(v))^\FF, ({%
	\bar{R}}_\alpha(u))^\FF]_\mathcal{F} \\[.3em]
&&[u^\FF ,[v^\FF,z^\FF]_\FF ]_\mathcal{F} =[[u^\FF,v^\FF]_\FF ,z^\FF]_\mathcal{F} + [({\bar{R}}%
^\alpha(v))^\FF, [({\bar{R}}_\alpha(u))^\FF,z^\FF]_\mathcal{F} ]_\mathcal{F} ~,\label{braid2}
\end{eqnarray}
for a proof we refer to eq. (3.7) (3.9), (3.10) in 
\cite{GR2}; indeed the braided Lie algebras presented here and there 
are isomorphic via $\DD^{-1}$ (cf. also \cite[\S 7 and \S 8]{LNP-book} and
\cite{ABP}).

In particular, infinitesimal translations $P^\mathcal{F}_\mu$ are given
by 
\begin{equation}
P_{\mu }^{\mathcal{F}}=\DD(P_\mu)=P_{\mu }\frac{1}{1+\frac{1}{\kappa
  }P_{0}}
~,\label{FF40}
\end{equation}
satisfy the braided Leibniz rule
\eqref{LeibFF}, that explicitly reads
$P_{\mu }^{\mathcal{F}}(f\st h)=P_{\mu }^{\mathcal{F}}(f)\st
h+e^{-{\sigma}}(f) \st P_{\mu }^{\mathcal{F}}(h)\,,
$
and have vanishing  braided
commutator, $[P_\mu^\FF,P_\nu^\FF]_\mathcal{F} =0$. 

In order to obtain \eqref{FF40}  use that
$\FF^{-1}=\exp(-iD\otimes -\sigma)$ and that $-iD$ on momenta acts as the identity operator:
$-iD(P_\mu)=[-iD,P_\mu]=P_\mu$, hence $\FF^{-1}=\exp(-iD\otimes
-\sigma)=\exp(1\otimes -\sigma)$ if the first leg of $\FF^{-1}$ acts
on $P_\mu$. The other expressions are computed in \cite[\S 3.2, 3.3]{ABP}.

As a consistency check of the differential geometry presented 
we can verify that in the equality 
$
\mathrm{d}f=\mathrm{d}x^\mu \star \partial_\mu^\FF f~,
$ 
obtained in \eqref{delF}
the quantum partial derivative $\partial_\mu^\FF
f=\frac{1}{1-\frac{i}{\kappa}\partial_0}\partial_\mu$ is exactly 
$\partial_\mu^\FF=\DD(\partial_\mu)$.
This confirms the interpretation of  $\partial_\mu^\FF=iP_\mu^\FF$ as infinitesimal translations.

\section{Group velocity in FLRW cosmology}\label{GVNC}
 We here briefly review the study of the wave
equation in 2 and 4 dimensions for a massless scalar field conformally coupled to 
the Friedman-Lemaitre-Robertson-Walker (FLRW) spacetime
(see e.g. \cite{Jacobson}, (\S 5.2), and \cite{MW} (\S 6.1)\/) and in
particular derive the group velocity $v_{g}$ of a wave packet in these
spacetimes. 
This study is propaedeutical to the noncommutative one in Section
\ref{DNCC}. The FLRW metric using
comoving coordinates is given by
\begin{equation}
\d s^{2}=-\mathrm{d}t^{2}+a^{2}\left( t\right) \mathrm{d}x^i\d x^i  \label{frwl}
\end{equation}%
where $a\left( t\right) $ is the cosmological expansion factor (scale
factor).
Correspondingly, the wave equation in 2 dimensions $\Box \varphi =0$ explicitly reads
\begin{equation}
\Box \varphi =(-\partial _{t}^{2}-a^{-1}\dot{a}\partial _{t}+a^{-2}\partial 
_{x}^2)\,\varphi =0  \label{FRWLWE}
\end{equation}
where $\dot{a}=\frac{\partial a}{\partial t}$. The solution of this linear 
differential equation can be found with the method of separation of 
variables, we set 
\begin{equation}
\varphi _{k}(x,t)=\lambda _{k}(t)\,e^{-i{k}{x}}=e^{i[f_{k}(t)-kx]}
\label{sparm}
\end{equation}%
and observe that $\varphi _{k}$ solves the wave equation if $\lambda _{{k}%
}(t)$ satisfies 
$(\partial _{t}^{2}+a^{-1}\dot{a}\partial _{t}+a^{-2}k^{2})\lambda _{k}(t)=0~.$
Under the change of variables $t\rightarrow \eta =\int \frac{1}{a}\mathrm{d}%
t $, so that $\mathrm{d}\eta =\frac{1}{a}\mathrm{d}t$ ($\eta $ is the 
so-called conformal time because the metric becomes conformally flat $%
g=a^{2}(-\mathrm{d}\eta ^{2}+\mathrm{d}x^{2})$), this latter equation 
becomes the usual harmonic oscillator equation $(\partial _{\eta 
}^{2}+k^{2})\lambda =0$, hence we have the forward travelling wave solution $%
\varphi _{k}(x,t)=e^{i[k_{{}}\eta (t)-kx]}$ (as well as the backward 
travelling solution $\tilde{\varphi}_{k}(x,t)=e^{-i[k_{{}}\eta (t)+kx]}$). 
The spacetime points of constant phase of this wave satisfy $k_{{}}\eta 
(t)-kx=${\it const.}, hence the phase velocity is $v_{p}=\frac{\mathrm{d}x}{\mathrm{%
d}t}=\frac{\mathrm{d}\eta (t)}{\mathrm{d}t}=\frac{1}{a}$. 
\\

A more physical measure of the field propagation speed is given by the group
velocity. We derive its expression for a wave packet that
is a superposition of waves of the kind (\ref{sparm}),   with general
time dependence $f_{k}(t)$; these are waves that are harmonic in space but
not necessarily in time. We hence consider the wave packet $\varphi =\frac{1%
}{2\pi }\int_{k-\delta k}^{k+\delta k}a_{\tilde{k}}\varphi _{\tilde{k}}%
\mathrm{d}{\tilde{k}}$ that is  peaked around a given value $k$,
i.e., $\delta k/{k}<<1$. As usual we rewrite the wave packet factorising the
wave $\varphi _{k}=e^{i[f_{k}(t)-kx]}$, so that
\begin{eqnarray}
\varphi &=&\frac{1}{2\pi }\int_{k-\delta k}^{k+\delta k}a_{\tilde{k}}e^{i[f_{%
\tilde{k}}(t)-{\tilde{k}}x]}\mathrm{d}\tilde{k}  \notag \\
&=&\frac{1}{2\pi }\Big(\int_{k-\delta k}^{k+\delta k}a_{\tilde{k}}e^{i[f_{%
\tilde{k}}(t)-f_{k}(t)-(\tilde{k}-k)x]\,}\mathrm{d}\tilde{k}\Big)%
e^{i[f_{k}(t)-kx]}  \notag \\
&\simeq &\frac{1}{2\pi }\Big(\int_{k-\delta k}^{k+\delta k}a_{\tilde{k}}e^{i(%
\frac{\partial f_{{k}}(t)}{\partial k}-x)(\tilde{k}-k)}\mathrm{d}%
\tilde{k}\Big)e^{i[f_{k}(t)-kx]}~.
\end{eqnarray}%
We observe that the integral has a phase that is slowly varying in space
with respect to the phase of the wave $e^{i[f_{k}(t)-kx]}$, i.e., the wave
packet $\varphi $ can be described as the wave $A_{k}(x,t)e^{i[f_{k}(t)-kx]}$
with the amplitude $A_{k}(x,t)=\frac{1}{2\pi }\int_{k-\delta
k}^{k+\delta k}a_{\tilde{k}}e^{i(\frac{\partial f_{\tilde{k}}(t)}{\partial k}%
-x)(\tilde{k}-k)}\mathrm{d}\tilde{k}$ that is slowly varying in space.
The points in spacetime where this amplitude is constant are determined by
the condition $\frac{\partial f_{{k}}(t)}{\partial k}-x=${\it const.}; it
then follows that the shape of the amplitude moves with the velocity
\begin{equation}\label{gv}
v_{g}=\frac{\partial x}{\partial t}=\frac{\partial }{\partial t}\frac{%
\partial f_k(t)}{\partial k}=\frac{\partial }{\partial k}\frac{\partial f_k(t)}{%
\partial t}~
\end{equation}%
that is by definition the group velocity of the wave packet $\varphi $
peaked around the wave number $k$.

\vskip .4cm We now proceed to compute the group velocity $v_g$ for waves
satisfying the wave equation (\ref{FRWLWE}) of 2-dimensional FLRW spacetime;
in this case $f_k(t)=k\eta$ and hence
\begin{equation}\label{drag}
v_g=\frac{1}{a}
\end{equation}
equals the phase velocity. The fact that this velocity does not equal the
speed of light $c$ but (inserting $c$ that was previously set equal to 1)
is $c/a$ should not be a surprise because the comoving reference frame is
not a free falling reference frame.  These results agree also with the
kinematic light cone condition $\mathrm{d} s^2=-\mathrm{d} t^2+a^2(t)%
\mathrm{d} x^2=0$.
\\

In 4 dimensions a massless scalar field conformally coupled to gravity satisfies
the wave equation 
\begin{equation}
(\Box -\frac{1}{6}R)\varphi=0~~~~\mbox{ i.e. }~~~~  (a\partial _{t}^{2}+3\dot{a}\partial _{t} -a^{-1}\partial_{x^i}^2+\frac{1}{6}aR)\,\varphi =0  \label{FRWLWE4}
\end{equation}
where $R$ is the scalar curvature. The solution of this linear differential equation can be found as before with the method of separation of variables by setting  
$
\varphi _{\bbk}(\boldsymbol{x},t)=\lambda _{\bbk}(t)\,e^{-i{\bbk}{\boldsymbol{x}}}
$ and then by considering conformal time $\eta(t)$. Recalling that
$R=6(\frac{\ddot{a}}{a^2}+\frac{\dot{a}^2}{a^2})=6\frac{a''}{a^3}$, the wave equation is equivalent to
$\lambda''_{\bbk}+2\frac{a'}{a}\lambda'_{\bbk}
+k^2\lambda_{\bbk}+\frac{a''}{a}\lambda_{\bbk}=0$, with
$k=\sqrt{k_x^2+k_y^2+k_z^2}$.
If we further set $\lambda_{\bbk}=a^{-1}\chi_{\bbk}$ the equation is solved iff
$\chi_{\bbk}$ satisfies the harmonic oscillator equation $\chi''+k^2\chi=0$.
Hence the wave equation \eqref{FRWLWE4} is solved by
$\varphi_{\bbk}(\boldsymbol{x},t)=a^{-1}e^{i(\omega \eta(t)-{\bbk}\boldsymbol{x})}$
with $\omega=k$.
\\

The group velocity for a superposition of waves of the kind
$\varphi_{\bbk}(\boldsymbol{x},t)=a^{-1}e^{i[f_{\bbk}(t)-{\bbk}\boldsymbol{x}]}$ 
is obtained, similarly to the 2 dimensional case,
by considering  a narrow wave packet centered around a 3-vector ${\bbk}$,
\begin{eqnarray}
\varphi &=&\frac{1}{(2\pi)^3 }\int_{{\bbk}-\boldsymbol{\delta} {\bbk}}^{{\bbk}+\boldsymbol{\delta} {\bbk}}a_{\tilde{\bbk}}a^{-1}e^{i[f_{%
\tilde{\bbk}}(t)-{\tilde{\bbk}}\boldsymbol{x}]}\mathrm{d}\tilde{\bbk}  \notag \\
&=&\frac{1}{(2\pi)^3}\Big(\int_{{\bbk}-\boldsymbol{\delta} {\bbk}}^{{\bbk}+\boldsymbol{\delta} {\bbk}}a_{\tilde{\bbk}}e^{i[f_{%
\tilde{\bbk}}(t)-f_{\bbk}(t)-(\tilde{\bbk}-{\bbk})\boldsymbol{x}]\,}\mathrm{d}\tilde{\bbk}\Big)%
a^{-1}e^{i[f_{\bbk}(t)-{\bbk}\boldsymbol{x}]}  \notag \\
&\simeq &\frac{1}{(2\pi)^3 }\Big(\int_{{\bbk}-\boldsymbol{\delta} k}^{{\bbk}+\boldsymbol{\delta} {\bbk}}a_{\tilde{\bbk}}e^{i(%
\nabla f_{{\bbk}}(t)-\boldsymbol{x})(\tilde{\bbk}-{\bbk})}\mathrm{d}%
\tilde{\bbk}\Big)a^{-1}e^{i[f_{\bbk}(t)-{\bbk}\boldsymbol{x}]}~.
\end{eqnarray}%
The slowly varying phase $A_{k}(x,t)=
\frac{a^{-1}}{(2\pi)^3 }\int_{{\bbk}-\boldsymbol{\delta} k}^{{\bbk}+\boldsymbol{\delta} {\bbk}}a_{\tilde{\bbk}}e^{i(%
\nabla f_{{\bbk}}(t)-\boldsymbol{x})(\tilde{\bbk}-{\bbk})}\mathrm{d}%
\tilde{\bbk}$ of this wave packet has modulus square
$A_{\bbk} A^{*_{}}_{\bbk}(\boldsymbol{x},t)= 
\frac{a^{-2}}{(2\pi)^6 }\int_{{\bbk}-\boldsymbol{\delta}
  k}^{{\bbk}+\boldsymbol{\delta} {\bbk}}
\int_{{\bbk-}\boldsymbol{\delta} k}^{{\bbk}+\boldsymbol{\delta}
  {\bbk}}a^{}_{\tilde{\bbk}} a^*_{\hat{\bbk}}
\cos\!\big((\nabla f_{{\bbk}}(t)-\boldsymbol{x})(\tilde{\bbk}-\hat{\bbk})\big)
\mathrm{d}\tilde{\bbk}\:\!\mathrm{d}\hat{\bbk}$ 
that is maximal when the cosine is maximal, hence for $\nabla
f_{{\bbk}}(t)-\boldsymbol{x}=0$, leading to the group velocity
$$
\boldsymbol{v}_g=\frac{\partial \boldsymbol{x}}{\partial
  t}=\frac{\partial}{\partial t} \nabla f_{\bbk}~.
$$ 

We have seen that massless scalar fields conformally coupled to 4 dimensional
FLRW-spacetime have waves $\varphi_{\bbk}(\boldsymbol{x},t)=a^{-1}e^{i(\omega \eta(t)-{\bbk}\boldsymbol{x})}$
with $\omega=k$ and hence the resulting group velocity has modulus
$v_g=\frac{c}{a}$ (we have restored the velocity of light
$c$). Considering a free falling reference frame, rather than a
comoving one, this becomes the velocity of light $c$ as expected.

\section*{Acknowledgements}
We would like to thank Antonaldo Diaferio and Attilio Ferrari for
useful discussions.  
This work is partially supported by a grant from Universit\`a Piemonte Orientale.
P.$\,$A. is partially supported from INFN, CSN4, Iniziativa
Specifica GSS.   
A.$\,$B. has been supported by the Polish National Science Center, project
2017/27/B/ST2/01902.
P.$\,$A. is affiliated to INdAM, GNFM (Istituto Nazionale di Alta Matematica, Gruppo Nazionale di Fisica Matematica).

\end{document}